\documentstyle[12pt,aasms4]{article}

\begin{document}
%
%
%
   \def\prd#1#2#3#4{#4 19#3 Phys.~Rev.~D,\/ #1, #2 }
   \def\pl#1#2#3#4{#4 19#3 Phys.~Lett.,\/ #1, #2 }
   \def\prl#1#2#3#4{#4 19#3 Phys.~Rev.~Lett.,\/ #1, #2 }
   \def\pr#1#2#3#4{#4 19#3 Phys.~Rev.,\/ #1, #2 }
   \def\prep#1#2#3#4{#4 19#3 Phys.~Rep.,\/ #1, #2 }
   \def\pfl#1#2#3#4{#4 19#3 Phys.~Fluids,\/ #1, #2 }
   \def\pps#1#2#3#4{#4 19#3 Proc.~Phys.~Soc.,\/ #1, #2 }
   \def\nucl#1#2#3#4{#4 19#3 Nucl.~Phys.,\/ #1, #2 }
   \def\mpl#1#2#3#4{#4 19#3 Mod.~Phys.~Lett.,\/ #1, #2 }
   \def\apj#1#2#3#4{#4 19#3 Ap.~J.,\/ #1, #2 }
   \def\aj#1#2#3#4{#4 19#3 Astr.~J.,\/ #1, #2}
   \def\acta#1#2#3#4{#4 19#3 Acta ~Astr.,\/ #1, #2}
   \def\rev#1#2#3#4{#4 19#3 Rev.~Mod.~Phys.,\/ #1, #2 }
   \def\nuovo#1#2#3#4{#4 19#3 Nuovo~Cimento~C,\/ #1, #2 }
   \def\jetp#1#2#3#4{#4 19#3 Sov.~Phys.~JETP,\/ #1, #2 }
   \def\sovast#1#2#3#4{#4 19#3 Sov.~Ast.~AJ,\/ #1, #2 }
   \def\pasj#1#2#3#4{#4 19#3 Pub.~Ast.~Soc.~Japan,\/ #1, #2 }
   \def\pasp#1#2#3#4{#4 19#3 Pub.~Ast.~Soc.~Pacific,\/ #1, #2 }
   \def\annphy#1#2#3#4{#4 19#3 Ann. Phys. (NY), \/ #1, #2 }
   \def\yad#1#2#3#4{#4 19#3 Yad. Fiz.,\/ #1, #2 }
   \def\sjnp#1#2#3#4{#4 19#3 Sov. J. Nucl. Phys.,\/ #1, #2 }
   \def\astap#1#2#3#4{#4 19#3 Ast. Ap.,\/ #1, #2 }
   \def\anrevaa#1#2#3#4{#4 19#3 Ann. Rev. Astr. Ap.,\/ #1, #2
                       }
   \def\mnras#1#2#3#4{#4 19#3 M.N.R.A.S.,\/ #1, #2 }
   \def\jdphysics#1#2#3#4{#4 19#3 J. de Physique,\/ #1,#2 }
   \def\jqsrt#1#2#3#4{#4 19#3 J. Quant. Spec. Rad. Transfer,\/ #1,#2 }
   \def\jetpl#1#2#3#4{#4 19#3 J.E.T.P. Lett.,\/ #1,#2 }
   \def\apjl#1#2#3#4{#4 19#3 Ap. J. (Letters).,\/ #1,#2 }
   \def\apjs#1#2#3#4{#4 19#3 Ap. J. (Supp.).,\/ #1,#2 }
   \def\apl#1#2#3#4{#4 19#3 Ap. Lett.,\/ #1,#2 }
   \def\astss#1#2#3#4{#4 19#3 Ap. Sp. Sci.,\/ #1,#2 }
   \def\nature#1#2#3#4{#4 19#3 Nature,\/ #1,#2 }
   \def\spscirev#1#2#3#4{#4 19#3 Sp. Sci. Rev.,\/ #1,#2 }
   \def\advspres#1#2#3#4{#4 19#3 Adv. Sp. Res.,\/ #1,#2 }
   %
%
%
\def\Msun{M_{\odot}}
\def\Mdot{\dot M}
\def\deg{$^\circ$\ }
\def\etal{{\it et~al.\ }}
\def\eg{{\it e.g.,\ }}
\def\etc{{\it etc.}}
\def\ie{{\it i.e.,}\ }
\def\ksec{{km~s$^{-1}$}}
\def\arcsec{{$^{\prime\prime}$}}
\def\arcmin{{$^{\prime}$}}
\def\subsun{_{\twelvesy\odot}}
\def\sun{\twelvesy\odot}
\def\gtwid{\mathrel{\raise.3ex\hbox{$>$\kern-.75em\lower1ex\hbox{$\sim$}}}}
\def\ltwid{\mathrel{\raise.3ex\hbox{$<$\kern-.75em\lower1ex\hbox{$\sim$}}}}
\def\plusminus{\mathrel{\raise.3ex\hbox{$+$\kern-.75em\lower1ex\hbox{$-$}}}}
\def\minusplus{\mathrel{\raise.3ex\hbox{$-$\kern-.75em\lower1ex\hbox{$+$}}}}
\title{Multi-Wavelength Variability of the Synchrotron Self-Compton Model for Blazar 
Emission}

\author{Mark W. Sincell}
\affil{Department of Physics MC 704 \\The University of Illinois at
Urbana-Champaign \\1110 W. Green Street \\Urbana, IL 61801-3080}
\authoraddr{Department of Physics MC 704 \\The University of Illinois at
Urbana-Champaign \\1110 W. Green Street \\Urbana, IL 61801-3080}

\begin{abstract}
Motivated by recent reports of strongly correlated radio and X-ray variability
in 3C279 (Grandi, etal 1995),
we have computed the relative amplitudes of variations in the synchrotron flux 
at
$\nu$ 
and the self-Compton X-ray flux at 1 keV ($R(\nu)$) for a homogeneous 
sphere of relativistic electrons orbiting in a tangled magnetic field.  
We consider three cases:
the Thomson depth of the sphere ($\tau_T$) varies at fixed magnetic field 
strength ($B$), $B$ varies at fixed $\tau_T$ and equal fractional changes in
$\tau_T$ and $B$.  
Relative to synchrotron self-Compton scattering without induced Compton
scattering, stimulated scattering reduces the amplitude of $R(\nu)$
by as much as an order of magnitude when $\tau_T \gtwid 1$.  
When $\tau_T$ varies in a fixed magnetic field, $R_{\tau}$ 
increases monotonically from 0.01 at $\nu_o$, the self-absorption turnover 
frequency, to $0.5$ at $100 \nu_o$.  Variations in $B$ increase $R$ at
all $\nu$, up to a factor of 2 if $\tau_T$ is constant,
and introduce local extrema in $R$.

The relative amplitudes of the correlated fluctuations in the radio-mm and X-ray
fluxes from 3C279 
are consistent with the synchrotron self-Compton model if $\tau_T$ varies in
a fixed magnetic field and induced Compton scattering is the dominant 
source of radio opacity.
The variation amplitudes are 
are too small to be produced by the passage of a shock through
the synchrotron emission region unless the magnetic field is 
perpendicular to the shock front.  
\end{abstract}

\section{Introduction}
Blazars are 
variable, polarized, 
flat spectrum extragalactic radio sources with a non-thermal continuum 
extending to $\gamma$-ray energies (\eg Urry \& Padovani 1995).
The radio emission from blazars is collimated into narrow beams composed of
many individual knots and an optically thick core (Kellerman \& Pauliny-T\"oth
1981).
In virtually all blazars, the radio knots appear to separate from the core
at speeds greater than the speed of light 
(Urry \& Padovani 1995)
and this superluminal motion is strong evidence that blazars are relativistic
jets of magnetized plasma viewed along the jet axis 
(Blandford \& K\"onigl 1979).
Although the jet model also accounts for such diverse blazar 
properties as
the flat radio spectrum (Kellerman \& Pauliny-T\"oth 1981), 
the short variability time scales
at all energies (\eg Quirrenbach, \etal 1991, Maraschi, \etal 1992), 
the small synchrotron self-Compton X-ray fluxes (Marscher 1987,
Ghisellini, \etal 1992) and the
large $\gamma$-ray fluxes (Maraschi, \etal 1992),
the radiative processes which produce the blazar continuum have not been
identified.  In this paper, we calculate the relative amplitudes of variations
in the radio and X-ray fluxes for 
one of the most popular models of the blazar continuum,
synchrotron self-Compton scattering in a relativistic outflow
(Jones, O'Dell \& Stein 1974, Marscher 1977,
K\"onigl
1981, Ghisellini, \etal 1985), and we compare the results to 
observations of correlated
variations on the radio and X-ray flux from 3C279 (Grandi, etal 1995: G95).

Blazar spectra are nearly featureless and 
a large number of models, which make very different assumptions for the
relevant radiation processes, agree qualitatively with
snapshots of the radio to $\gamma$-ray spectrum (Maraschi, \etal 1992,
Maraschi, \etal 1994,
Hartmann \etal 1996).  At low energies, the flat radio spectrum of compact
radio cores can be explained by inhomogeneous synchrotron emission 
(Marscher 1977, K\"onigl
1981, Ghisellini, \etal 1985) or by a homogeneous core that is optically thick 
to induced
Compton scattering (Sincell \& Krolik 1994: SK94).  The situation at high
energies is even more complicated.  The X- and $\gamma$-ray emission may
be synchrotron emission by high energy electrons (K\"onigl 1981,
Ghisellini, \etal 1985), radiation
from a pair cascade (Blandford \& Levinson 1995) or low frequency photons which are inverse
Compton scattered by relativistic electrons in the jet.
In the last case, the source
of the low energy photons could be synchrotron radiation from the jet (the
synchrotron self-Compton model, Maraschi \etal 1992, 
Maraschi, \etal 1994, Bloom \& Marscher 1996), UV radiation from
a disk (Dermer, Schlickeiser \& Mastichiadis 1992), or some diffuse source of 
radiation surrounding the jet
(Sikora, Begelman \& Rees 1994, Ghisellini \& Madau 1996).
We will consider only synchrotron self-Compton scattering in this paper and
ignore other sources for the high energy emission.

The synchrotron self-Compton model for the continuum emission 
assumes that
a single population of relativistic electrons radiates synchrotron photons and
subsequently scatters a fraction of these to higher energies 
(Jones, \etal 1974).
Fluctuations in either the electron density, the magnetic field strength or
the Doppler factor of the emission region will affect the synchrotron and
self-Compton fluxes instantaneously.  Therefore, variations in the low and
high energy fluxes which are uncorrelated, or have significant time delays,
cannot be the result of synchrotron self-Compton scattering.  The relative
amplitudes of the variations in the synchrotron and self-Compton fluxes depends
upon which physical parameters change.  For example, the fractional change in
the high energy flux is twice as large as the fractional change in the
optically thin synchrotron flux
when the electron density varies, whereas they are equal if the magnetic field
or the Doppler factor changes (SK94).

The amplitude of the variations in the synchrotron and self-Compton fluxes
also depends upon the relativistic electron distribution.  The electron
spectrum will flatten with increasing radiation intensity because both the
synchrotron and inverse Compton cooling rates increase with increasing
radiation energy density.  
However, these cooling processes are both too slow to have any
effect upon the electron distribution in the parsec scale jet (SK94).  
Therefore,
we neglect the evolution of the electron spectrum in the calculations
presented in this paper.

SK94 demonstrated that induced Compton scattering can reduce the 
amplitudes of variations in the 
synchrotron radiation when the brightness temperature of the source 
$T_B \ltwid 2 \times 10^{11} \mbox{K}$.  
The amplitudes of variations in the self-Compton X-ray flux are unaffected by 
induced Compton scattering because the X-ray flux is dominated by 
photons scattered from
the high-frequency, low $T_B$, end of the synchrotron spectrum, where
the stimulated scattering optical depth is small.
Thus, the relative amplitudes of the variations in the
synchrotron flux at the self-absorption turnover frequency and the self-Compton
flux at 1 keV are reduced from $\sim 0.4$ to $\sim 0.2$ when the induced
Compton scattering optical depth is large (SK94).

The continuum emission from 3C279, one of the most intensively monitored
blazars, varies coherently over its entire spectrum (Maraschi, \etal 1992,
Maraschi \etal 1994, Hartmann, \etal 1996).
Recently, G95 used historical light curves to show that the
radio-mm and 1 keV X-ray fluxes from 3C279 are strongly correlated.
The maximum time resolution of the 3C279 light curves was $\sim 70 \mbox{d}$ 
and the absence of any detectable time delay 
implies that the radio-mm and X-ray fluxes
are from physically related regions separated by $\ltwid 0.06 \mbox{pc}$.
While this is consistent with the assumptions of the synchrotron self-Compton
theory, the relative amplitudes of the radio and X-ray variations are smaller
than predicted (Sincell 1996).  

In this paper, we extend previous calculations (SK94, Sincell 1996)
and compute the relative amplitudes of 
variations in the
synchrotron self-Compton flux as a function of the frequency of 
the synchrotron emission. We first define the flux variability ratio (\S
\ref{sec: flux variability ratio}) and describe how it is calculated.  This
ratio is computed for three simple types of variations in \S\ref{sec: results}
and the implications for 3C279 are discussed in \S \ref{sec: 3C279}.  We
conclude in \S \ref{sec: conclusions}.

\section{\bf The Flux Variability Ratio}
\label{sec: flux variability ratio}
The time variability of the source spectrum is approximated as a sequence of 
steady
state spectra.
We have used the code developed in SK94 to compute the steady-state
synchrotron spectrum and
self-Compton X-ray flux from a homogeneous sphere containing an isotropic
power-law distribution of relativistic electrons
\begin{equation}
\label{eq: electron distribution}
{\partial n_e \over \partial \gamma} = n_o \gamma^{-p}
\end{equation}
for $\gamma \geq \sqrt{2}$. We assume $p=2.5$ in all the simulations and the 
normalization ($n_o$) is fixed by the assumed $\tau_T$.
This code incorporates synchrotron absorption and emission, inverse Compton
scattering and induced Compton scattering by the relativistic electrons.
Stimulated scattering becomes the dominant source of radio opacity when 
$T_B \gtwid 2 \times 10^{11} \mbox{K}$ (SK94) 
and must be included when calculating the spectra of compact radio sources.
Self-absorption reemerges as the dominant opacity source at low frequencies
and inverse Compton scattering of synchrotron photons by electrons with
$\gamma \ltwid 10$ contributes to the radio flux above the self-absorption
turnover.

The flux variability ratio 
\begin{equation}
R_m(\tau_T,B,\nu) = {\partial \log S_r(\tau_T,B,\nu) 
      \over \partial \log S_x(\tau_T,B)}|_m,
\end{equation}
is the ratio of the fractional change in the synchrotron flux at 
$\nu$ ($S_r$) and
the self-Compton X-ray flux ($S_x$) at 1 keV caused by a fluctuation 
in the physical 
parameter $m$.  In this paper we investigate three different variations:
the Thomson
depth ($\tau_T$) of the source varies at fixed magnetic field strength 
($m=\tau$), 
the magnetic field strength ($B$) varies at fixed $\tau_T$ ($m=B$) and equal
fractional changes in $\tau_T$ and $B$ $(m=S)$.    
The third case approximates the 
passage of a strong shock through the plasma, assuming that the field is 
tangled (\eg Marscher \& Gear 1985).
We also assume that $\tau_T$ and $B$ are uniform throughout the source.  
Simple analytic calculations of $R_m$ are possible for 
synchrotron self-Compton scattering (SK94),
but $R_m$ must be calculated numerically when the induced Compton 
scattering opacity is large (SK94).

We compute the variability ratio using the approximate formula
\begin{equation}
R_m(\nu) \simeq {S_r(m,\nu) - S_r(m+\Delta m,\nu) \over
                 S_r(m,\nu) + S_r(m+\Delta m,\nu)}
         \cdot  {S_x(m)     + S_x(m+\Delta m)     \over
                 S_x(m)     - S_x(m+\Delta m)},
\end{equation}
and the radio spectra and X-ray fluxes from two models 
with closely spaced values of 
the parameter
$m$.  In this paper we choose $\Delta m/m =0.1$ but the results are fairly
insensitive to $\Delta m$, even when $\Delta m/m \gtwid 1$ (see figures in
Sincell 1996).

\section{Results}
\label{sec: results}

We have calculated $R_{\tau}$, $R_{B}$  and $R_S$ for 
$0.01 \leq \tau_T \leq 3.0$
and a range of $B$.
In figs. \ref{fig: tau variability}, \ref{fig: b variability}
and \ref{fig: shock variability}  we plot $R_m$ as a function of $\nu / \nu_o$,
where $\nu_o$ is the synchrotron self-absorption turnover 
frequency.  
When the effects of induced and inverse Compton scattering on the radio
spectrum are neglected, $R_m(\nu/\nu_o)$ is independent of
the unperturbed values of $\tau_T$ and $B$.  Including Compton scattering
in the computation of the radio spectrum
introduces a strong dependence upon $\tau_T$ but $R_m(\nu/\nu_o)$ remains
nearly independent of $B$ because the induced Compton scattering optical depth
at $\nu_o$ is $\propto T_B \tau_T \propto B^{-1/(p+4)} \tau_T^{(p+5)/(p+4)}$
(SK94).
The code was used to verify that $R_m$ changed by less than a few percent when
$B$ increased by two orders of magnitude.
Therefore, we present $R_m$ for $B=10^{-5} \mbox{G}$
and use the relation (\eg SK94)
\begin{equation}
\label{eq: nuo scaling}
\nu_o \propto
\left( {\delta \over 1+z} \right)
B^{(p+2)/(p+4)} 
\end{equation}
to scale $R_m$ to any desired field strength, 
Doppler boost ($\delta$) or redshift ($z$).

Although $R_m(\nu/\nu_o)$ is 
independent of the unperturbed values of $\tau_T$ and $B$ when the effects
of Compton 
scattering on the radio spectrum are neglected, 
it does depend on which parameters vary. 
At optically thin frequencies, $\nu \gg \nu_o$,
$R_{\tau} \simeq 0.5$, $R_{B} \simeq 1.0$ and $R_{S} \simeq 0.7$, 
independent of frequency  
(figs. \ref{fig: tau variability}, \ref{fig: b variability}
and \ref{fig: shock variability}).   
These numerical values are in good agreement with the
analytic results in SK94.

Synchrotron self-absorption reduces $R_m$ at $\nu \ltwid \nu_o$.
This is because any
change in the physical parameters which increases the synchrotron
emissivity results
in a compensating increase in the opacity and decrease in the photospheric 
depth.
Thus, the net flux at optically thick frequencies is less 
sensitive to changes in the source parameters.
The ratio of the self-absorption opacity to the emissivity
increases rapidly as $\nu/\nu_o$ decreases (\eg SK94)
and
$R_{\tau}$ approaches zero  at $\nu \ltwid
\nu_o / 2$.  
Increasing $B$ also increases $\nu_o$ (eq. \ref{eq: nuo scaling}) and
the increase in the self-absorption opacity at fixed $\nu$
overwhelms the increase in the
emissivity when $\nu \ll \nu_o$.
The resulting decrease in the synchrotron flux at $\nu$ appears as
negative values of $R_{B}$ and $R_{S}$
(figs. \ref{fig: b variability}
and \ref{fig: shock variability}). 
Negative values of $R_m$ correspond to an anti-correlation
of the synchrotron and self-Compton fluxes.   

Compton scattering changes the frequency dependence
of $R_m$ and introduces a dependence on $\tau_T$ 
(figs. \ref{fig: tau variability}, \ref{fig: b variability}
and \ref{fig: shock variability}).
Induced Compton scattering reduces $R_{\tau}$ over more than
a decade in frequency when $\tau_T \gtwid 0.1$ 
(fig. \ref{fig: tau variability}), relative to the 
synchrotron self-Compton scattering model without stimulated scattering. 
$R_{\tau}$ at $\nu \simeq \nu_o$ is reduced by almost
an order of magnitude when $\tau_T \gtwid 1$.
The stimulated scattering opacity at low frequencies and the 
contribution of inverse Compton scattered photons at higher frequencies
results in a monotonic increase in $R_{\tau}$ from $0.01$ at $\nu \ltwid 
\nu_o$ to $0.5$ at $\nu \sim 100 \nu_o$ (fig. \ref{fig: tau variability}).  

Even though synchrotron self-absorption is the dominant source of opacity, 
stimulated scattering increases the photon 
occupation numbers at 
low frequencies.  This increases the synchrotron flux at $\nu \ll \nu_o$
and the 
anti-correlation of the synchrotron and self-Compton fluxes caused
by variations in $B$ disappears when $\tau_T \gtwid 1$.

Variations
in the magnetic field strength introduce local extrema into $R_{B,S}$ when
the stimulated scattering optical depth is large.
The largest contribution
to the induced Compton scattering opacity at $\nu \ltwid \nu_o$ is from 
electrons 
with 
$\gamma_{*} 
= {1 \over 2} \left( {\nu_m \over \nu} \right)^{1/2}$
where $\nu_m \gtwid \nu_o$ is the peak of the spectrum (SK94). 
The low energy cutoff $\gamma = \sqrt{2}$ reduces the stimulated scattering 
opacity at frequencies
$ \nu_m/8 \ltwid \nu \ltwid 8\nu_m$
because $\gamma_* < \sqrt{2}$ and there are no electrons which couple 
$\nu$ to $\nu_m$.
When synchrotron self-absorption is the dominant source of opacity, 
the optical depth of the plasma
decreases with frequency and $R$ increases with frequency.  
A local maximum in $R$ appears at the frequency where the stimulated 
scattering opacity is approximately equal to the synchrotron opacity.  At 
higher frequencies, induced Compton scattering
limits the variations in the synchrotron flux and reduces $R$.
The local minimum in $R$ occurs at $\nu \sim \nu_m/8$ where the stimulated
scattering opacity
reaches a maximum.

These local extrema are not as prominent in $R_{\tau}$ because variations
in $\tau_T$ increase the optical depth at all frequencies.
However, the kink in the $\tau_T = 0.1$ curve of $R_{\tau}$ 
(fig. \ref{fig: tau variability}) is also due to this effect.

\section{\bf 3C279}
\label{sec: 3C279}
G95 used historical light curves of 3C279 to show that variations in
the radio-mm and
X-ray fluxes are strongly correlated with a time delay of $\ltwid 70$~days.
They also calculated the logarhithmic dispersion, or variability amplitude 
($v(\nu)$), of
the available measurements and found that $v(\nu)$ increases systematically 
with frequency.
However, these estimates of $v(\nu)$ are uncertain because many emission
components contribute to the observed flux at a given frequency (\eg
Unwin, \etal 1989).  These components may vary independently and the G95
data lacks the spatial resolution necessary to reliably subtract the
non-variable background.
Long-term monitoring at higher resolution (VLBA) is necessary to remove this
source of uncertainty.  In the remainder of this paper we will assume that
the variable component dominates the observed flux, but it should be remembered
that a significant amount of non-variable flux at $\nu$ will reduce $v(\nu)$ 
below the value expected for the variable component alone.

Both the strong correlation of the radio and X-ray fluxes and the absence
of a detectable time delay between variations in the two bands 
are consistent
with the assumptions of the synchrotron self-Compton model for the continuum
emission.
We used the variability amplitudes calculated by G95
to estimate 
\begin{equation}
R \simeq {v(\nu) \over v(1 \mbox{keV})}
\end{equation}
at four frequencies in the range $14.5 \mbox{GHz} < \nu < 230 \mbox{GHz}$.
The estimated $R$ for the epochs 1988-1991.4 and 1991.4-1993.2
are plotted in fig. \ref{fig: observed variability}.  
The model $R_{\tau}$ for $\tau_T = 1.0$, $B=10^{-3} \mbox{G}$ and
$\delta = 20$ is displayed on the same figure.

The relative amplitudes of the variations in the radio and X-ray fluxes from
3C279 are consistent with the synchrotron self-Compton model if $\tau_T$
varies in a fixed magnetic field and induced Compton scattering is the 
dominant source of radio opacity.
It is immediately apparent from fig. \ref{fig: observed variability}
that the magnitudes of both $R_{B}$ and $R_{S}$ are
too large to fit the observed values of $R$ for 3C279.
In addition, neither the local extrema in $R$ or the anti-correlation
of the synchrotron and self-Compton fluxes are observed.
We also find that the increase in $R$ with
frequency is much slower than expected for the synchrotron self-Compton
model without induced Compton scattering
(fig. \ref{fig: observed variability}).   However, both the magnitude
and the frequency dependence of $R$ are consistent with $R_{\tau}$ when
$\tau_T \gtwid 1$.  

This implies that $\tau_T
\sim 1$ in the synchrotron self-Compton emission region.  
The magnetic field strength and Doppler factor cannot be
calculated independently (eq. \ref{eq: nuo scaling}), but the values we have
assumed ($B = 10^{-3} \mbox{G}$ and $\delta = 20$) are consistent with
other estimates of the physical parameters for 3C279 (Ghisellini, etal 1985,
Maraschi, \etal 1992, SK94).  Larger magnetic fields require smaller Doppler
factors and vice versa.

We can set a lower limit on the electron density using the variability time
scale and the requirement $\tau_T \sim 1$.  The linear dimension of the
emission region
$l \ltwid l_{max} = 1.8 \times 10^{17} \delta \mbox{cm}$
if the observed flux varies on a time scale of $\ltwid 70$ days.  
The stimulated scattering optical depth of the plasma will be large enough to 
reduce the variability
amplitudes if the electron density 
$n_e \gtwid 8 \times 10^6 \delta^{-1} \mbox{ ${\rm cm^{-3}}$}$.
The total particle energy density of the distribution in eq. 
\ref{eq: electron distribution} is dominated by the rest mass  energy so
$U_e \sim n_e m_p c^2 \sim 7 \delta^{-1} \left( {m_p \over m_e} \right)
\mbox{ergs ${\rm cm^{-3}}$}$,
where $m_{p,e}$ are the masses of the positively charged particle and an
electron, respectively.  The magnetic energy
density $U_B \ll U_e$ and the plasma is far from equipartition.

A strong shock passing through the synchrotron emission region amplifies
both the electron density and the magnetic field strength if the magnetic
field is either tangled or aligned parallel to the shock front (\eg
Marscher \& Gear 1985).  Our results
for 3C279 indicate that the variations in the flux are due to fluctuations
in the electron density alone.  Thus, we conclude that if the variations
are caused by a shock the magnetic field must be aligned perpendicular to the
shock front.  An alternative possibility is that the observed variations are 
due to fluctuations in the local
electron density caused by a change in the particle injection rate.

\section{\bf Conclusions}
\label{sec: conclusions}

We have calculated $R_m(\nu)$, the relative amplitude of variations in the
synchrotron flux at
$\nu$ and the self-Compton
X-ray flux at 1 keV, for a homogeneous  sphere
of relativistic electrons orbiting in a tangled magnetic field.
The index $m$ refers to the physical quantity which is assumed to vary and
in this paper we investigate three cases: variations in $\tau_T$ at fixed
$B$, variations in $B$ at fixed $\tau_T$ and equal fractional changes in both
quantities.  The last case approximates the passage of a strong shock through
the plasma (\eg Marscher \& Gear 1985).  We find $R_m$ to be useful for two 
reasons.  First, the frequency dependence of $R_m$
is determined by the optical depth of the plasma.  Second, $R$ may be estimated
directly from observations of correlated radio and X-ray variability (\eg
G95). 

If synchrotron self-absorption is the dominant source of
opacity,  
the frequency dependence of $R_m$ is determined by $\nu_o$, the self-absorption
turnover frequency, and the physical parameter which is assumed to vary. 
We find that $R_m$ is constant at all optically thin
frequencies ($\nu \gg \nu_o$) and, for our assumed electron distribution 
(eq. \ref{eq: electron distribution}), $R_{\tau} \simeq 0.5$, $R_B \simeq 1.0$
and $R_S \simeq 0.7$.  Self-absorption causes all the $R_m$ to drop sharply
at $\nu \ltwid \nu_o$ and both $R_{B,S}$ become negative at $\nu \ll \nu_o$.

Induced Compton scattering reduces $R_m$ over more than a decade in
frequency, relative to the synchrotron self-Compton model without
stimulated scattering, when $\tau_T \gtwid 0.1$.
Increasing $\tau_T$ reduces $R_{\tau}$ at frequencies near $\nu_o$ 
and $R_{\tau}$
can be an order of magnitude smaller than the self-absorbed value when $\tau_T
\gtwid 1$.  We find that $R_{\tau}$ increases monotonically from low to high
frequencies, but a slight change in the stimulated scattering opacity at
$\nu \sim \nu_m / 8$ causes
local extrema in $R_{B,S}$.

Variations in the Thomson depth of a homogeneous source of synchrotron 
self-Compton 
radiation reproduces
the relative amplitudes of the correlated radio and X-ray flux variations in
3C279 (G95) 
if $\tau_T \sim 1$ and the emission region is optically
thick to induced Compton scattering.  Although $B$ and $\delta$ cannot be 
independently constrained (eq. \ref{eq: nuo scaling}), the observed $R$ is 
consistent with $B \sim 10^{-3} \mbox{G}$ and $\delta \sim 20$.  If we
assume that the maximum linear dimension of the emission region is $l_{max}
\sim 0.06 \delta$ pc, as implied by the variability time scale, $\tau_T \sim 1$
requires that the electron
density be $n_e \gtwid 8 \times 10^6 \delta^{-1}$~cm$^{-3}$.   In this case,
the particle energy density is much larger than the magnetic field energy
density.  

Variations in the magnetic field strength result in values
of $R$ which are larger than observed.  If the observed fluctuations are due
to the passage of a shock, the magnetic field must be oriented perpendicular
to the shock front.  Variations in the local particle injection rate could
change the electron density without necessarily changing the field strength.

Finally, it has been argued that stimulated scattering cannot be important in
blazars with optically thick spectral indices of $\alpha = -5/2$ 
(Litchfield, \etal 1995).  This 
argument is erroneous because synchrotron self-absorption {\it always} 
becomes the 
dominant source of opacity, and $\alpha = -5/2$, at low enough frequencies
(SK94).  However, this points out the ambiguities inherent in attempting to
determine the 
radio
opacity using spectral measurements alone.  Additional multi-wavelength 
variability
studies (\eg G95) are clearly necessary to determine the relevant
radiative processes in blazars. 

\acknowledgements
We thank the referee, Laura Maraschi, for helpful comments and Stephen Hardy 
for pointing out the error in the kinetic equation.
Support for this work was provided by 
NASA
grants NAGW-1583 and NAG 5-2925, and NSF grant AST 93-15133.  The simulations
were performed at the Pittsburgh Supercomputing Center (grant AST 960002P).

\begin{appendix}

\section{Correction to the Kinetic Equation}
There is an error in the SK94 expression for the contribution of 
induced Compton 
scattering to the photon 
kinetic equation.  
SK94 transformed the Eulerian
time derivative of the photon occupation number to the electron rest frame. 
However,
the correct method is to transform the Lagrangean time derivative 
(see SK94 eq. 15,
and Hardy \& Melrose 1994).
When this mistake is corrected we find that the expression for the mean rate of
change in the photon occupation number caused by stimulated scattering reads
\begin{eqnarray}
\langle {Dy(\nu) \over Dt}\rangle & = & {3 
\over 128
\pi^2} (\sigma_T c) y(\nu) {\partial y \over \partial\nu} \int {d\Omega_e \over
4\pi} \int d\gamma \left\{  {\partial n_e \over \partial\gamma} \right\} \gamma^{-3} \nonumber
\\ 
& & \times \int d\Omega^{\prime}_1 \int d\Omega^{\prime}_2
(1+\beta\mu^{\prime}_1)^{\alpha-1} (1+\beta\mu^{\prime}_2)^{-(\alpha+3)}
\left\{ 1+ (\hat k^{\prime}_1\cdot\hat k^{\prime}_2)^2 \right\} (\hat
k^{\prime}_1\cdot\hat b^{\prime} - \hat k^{\prime}_2\cdot \hat
b^{\prime})^2, 
\end{eqnarray}
where $D/Dt = \partial /\partial t + \bf\Omega_1 \cdot \bf\nabla$ 
is the Lagrangean 
derivative and the other variables are 
defined in SK94.  

This correction results in two small changes in the formulae of SK94.  First,
equation 6 should read
\begin{equation}
P(\gamma) = \cases{ (2\gamma)^{-(2\alpha + 5)} \left\{ {1 \over
(\alpha-1)^2} + {2 \over \alpha (\alpha+1) (\alpha-1)} \right\} & when
$\alpha < -1$; \cr (2\gamma)^{2\alpha -1} \left\{ {1 \over (\alpha+1)^2}
- {2 \over (\alpha+1)(\alpha+2) (\alpha+3)}\right\} & when $\alpha >
-1$ \cr}.
\end{equation}
This change does not affect any of the conclusions in SK94.
Second, equation 18 should read
\begin{equation}
\left[{Dy(\nu_1) \over D \tilde t}\right]|_{ics} 
= {1 \over 12\pi} y(\nu_1) \int
{d\Omega_e \over 4\pi} \int d\Omega_2 \int {d\nu_2 \over \nu_1} \gamma_o^{-p}
{\beta_o^2 (1 - \hat k_1 \cdot \hat k_2) \over |D_1 - D_2|}
{\partial y(\nu_2) \over \partial \nu_2}. 
\end{equation}
The change in the term in the kinetic equation results in an increase in the
peak synchroton flux (and $T_B$) of $\ltwid 7\%$, which is negligible.
\end{appendix}

\vfil\eject
\centerline{\bf Figure Captions}
\bigskip

\begin{figure}
\plotone{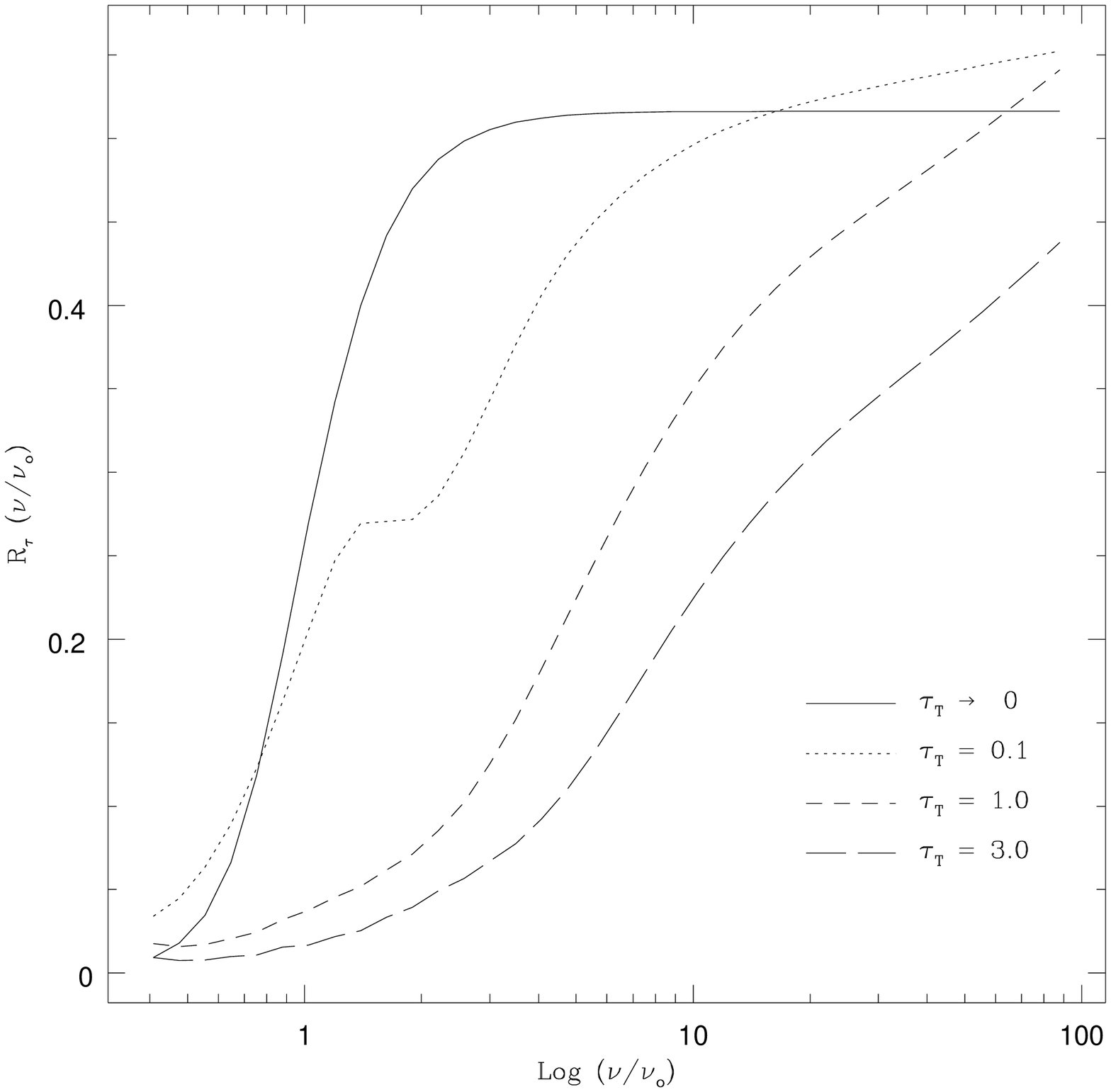}
\caption{\label{fig: tau variability}   }
\end{figure}

Figure \ref{fig: tau variability}. 
Relative variation in the radio flux density for variations in the Thomson
depth as a function of the total Thomson depth of the source.  The assumed
magnetic field strength is $B=10^{-5} \mbox{G}$.
The solid line, 
labeled $\tau_T \rightarrow 0$, corresponds 
to synchrotron
self-Compton scattering with neither inverse Compton nor induced Compton 
scattering included.

\begin{figure}
\plotone{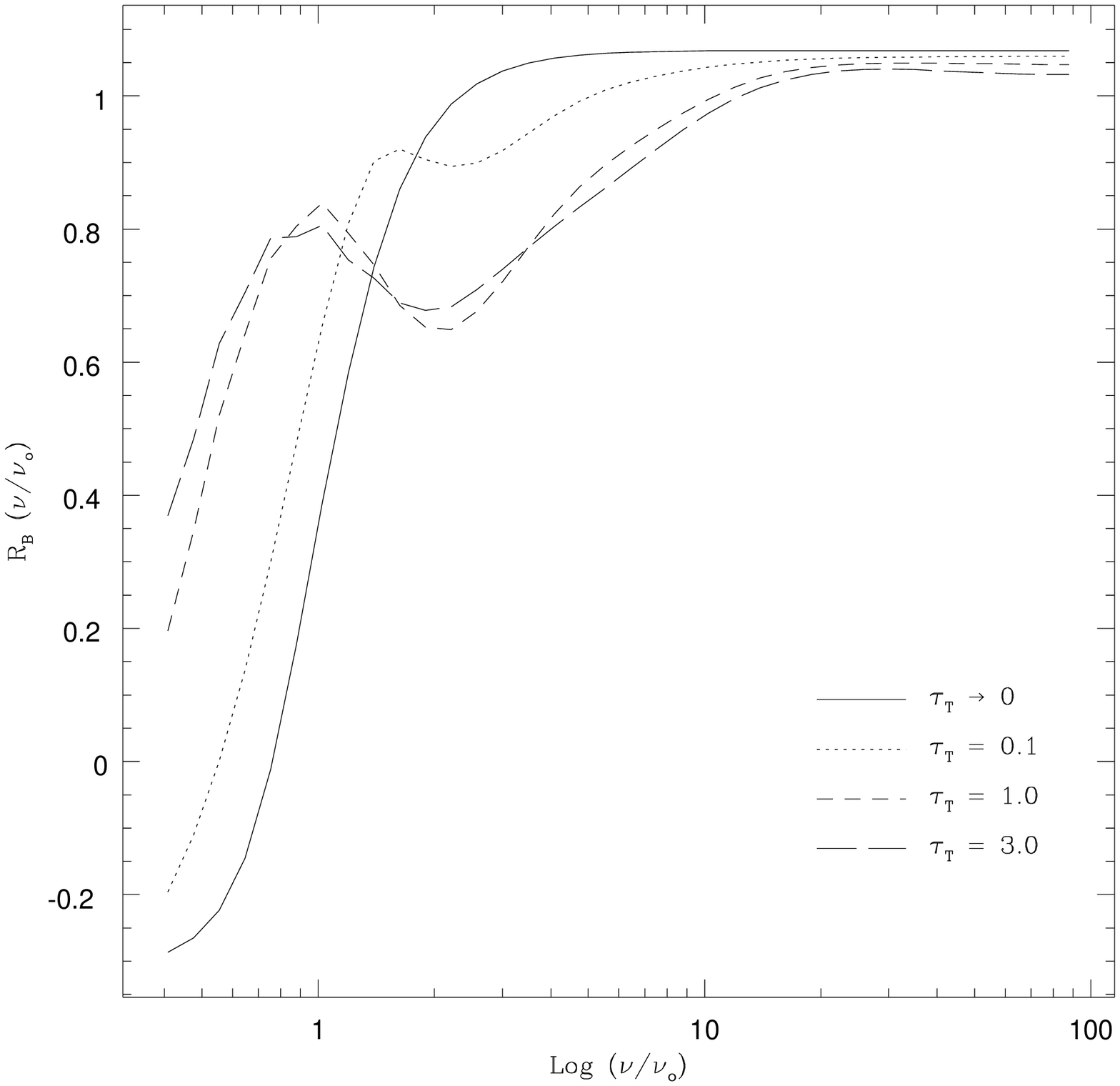}
\caption{\label{fig: b variability}   }
\end{figure}

Figure \ref{fig: b variability}. 
Relative variation in the radio flux density for variations in the magnetic
field as a function of the total Thomson depth of the source.  The assumed
magnetic field strength is $B=10^{-5} \mbox{G}$.
The solid line, 
labeled $\tau_T \rightarrow 0$, corresponds 
to synchrotron
self-Compton scattering with neither inverse Compton nor induced Compton 
scattering included.

\begin{figure}
\plotone{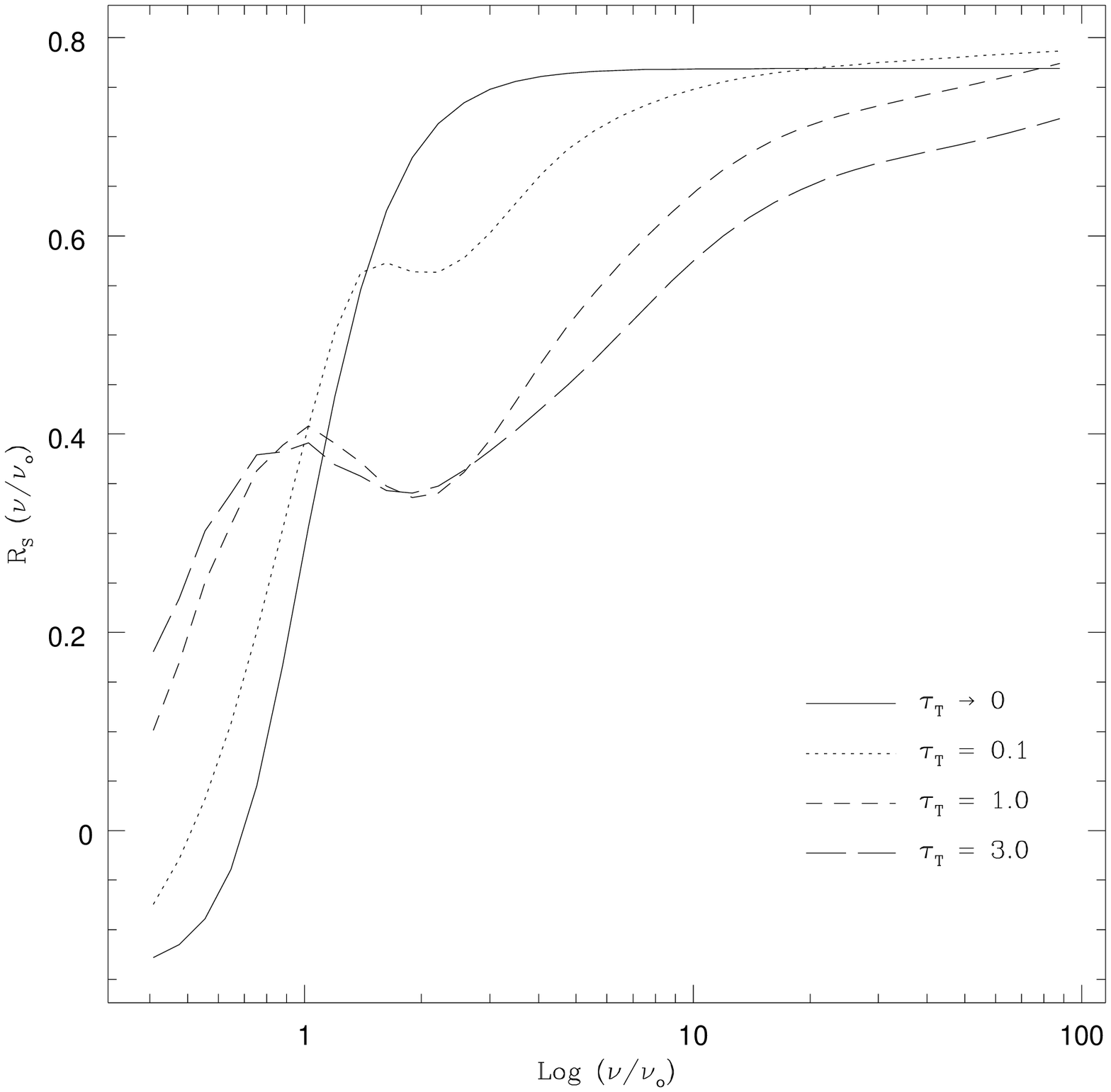}
\caption{\label{fig: shock variability}   }
\end{figure}

Figure \ref{fig: shock variability}. 
Relative variation in the radio flux density for equal amplitude variations in 
the Thomson depth and magnetic field strength
as a function of the total Thomson depth of the source.  
This is meant to mimic the variations caused by the passage of a shock through
the plasma.
The solid line, 
labeled $\tau_T \rightarrow 0$, corresponds 
to synchrotron
self-Compton scattering with neither inverse Compton nor induced Compton 
scattering included.

\begin{figure}
\plotone{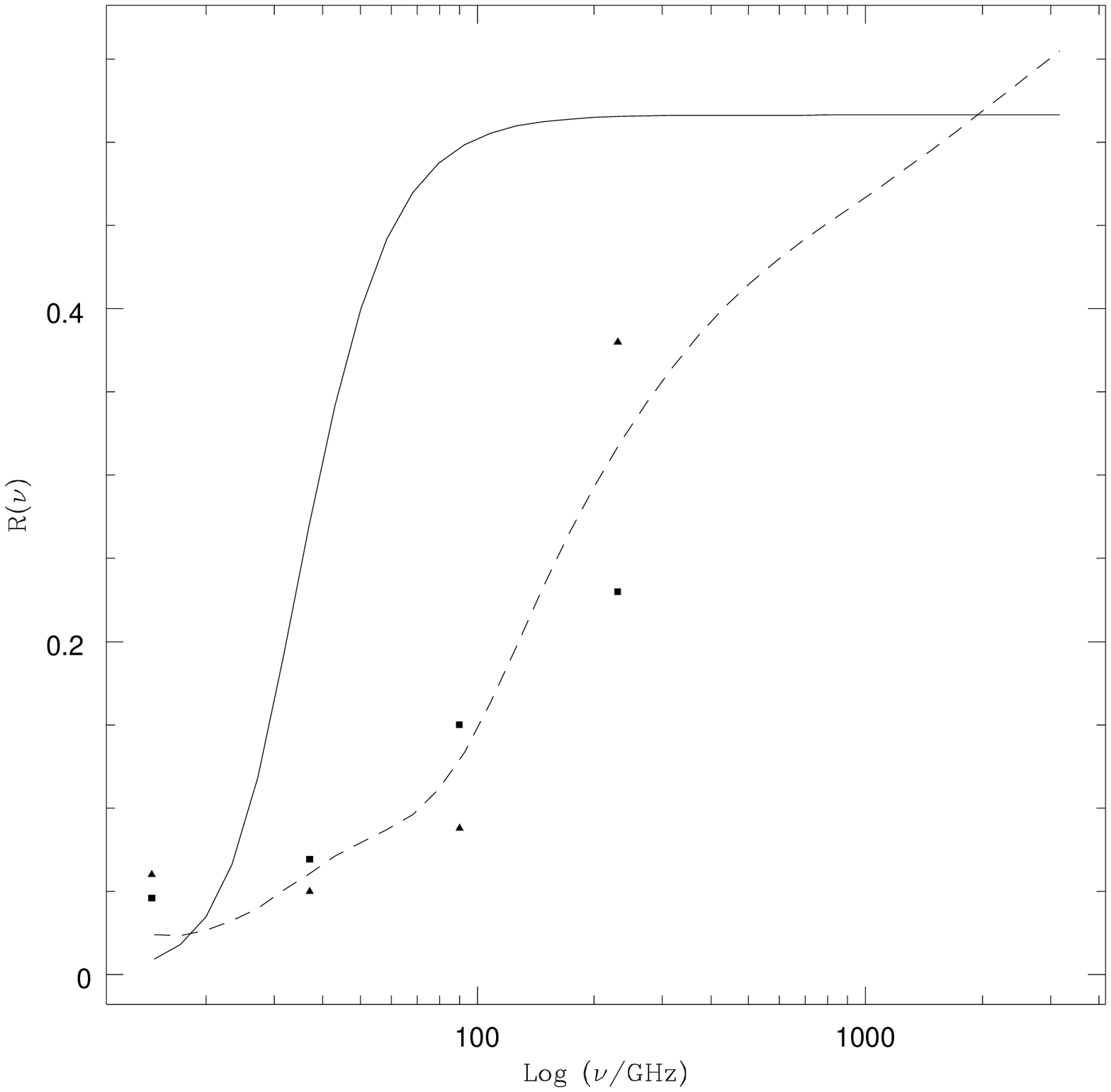}
\caption{\label{fig: observed variability}   }
\end{figure}

Figure \ref{fig: observed variability}. 
Comparison to the variability amplitudes
of 3C279 for the epochs 1988-1991.4 (squares) and 1991.4-1993.2 (triangles).
The solid line is the predicted amplitude for synchrotron
self-Compton scattering and the dashed line includes induced Compton scattering.
The magnetic field strength is $B=10^{-3} \mbox{G}$, the Thomson depth is
$\tau_{T} = 1.0$ and the Doppler factor
is $\delta = 20$.  

\end{document}